\documentclass[twocolumn]{aa}
\usepackage{graphicx}
\begin{document}
   \title{Simulations of thermally broadened H\,{\sc i} Ly\,$\alpha$ absorption arising 
          in the warm-hot intergalactic medium}


   \author{P. Richter,
          \inst{1}
          T. Fang,
          \inst{2}
	  \fnmsep\thanks{{\it Chandra} Fellow}
	  \and
	  G.L. Bryan
          \inst{3}
          }

   \offprints{P. Richter\\
   \email{prichter@astro.uni-bonn.de}}

   \institute{Argelander-Institut f\"ur Astronomie
              \thanks{Founded by merging of the Institut f\"ur Astrophysik
	      und Extraterrestrische Forschung, the Sternwarte, and the
	      Radioastronomisches Institut der Universit\"at Bonn.},
	      Universit\"at Bonn, Auf dem H\"ugel 71, 53121 Bonn, Germany
         \and
             Department of Astronomy, University of California at Berkeley, 
	     601 Campbell Hall, Berkeley, CA 94720, USA
         \and
	     Department of Physics, University of Oxford, Keble Road, Oxford OX13RH, UK
            }

   \date{Received xxx; accepted xxx}

\abstract{
Recent far-ultraviolet (FUV) absorption line 
measurements of low-redshift quasars have unveiled a population
of intervening broad H\,{\sc i} Ly\,$\alpha$ absorbers (BLAs)
with large Doppler parameters ($b\geq 40$ km\,s$^{-1}$). If
the large width of these lines is dominated by thermal line
broadening, the BLAs may trace highly-ionized gas in the 
warm-hot intergalactic medium (WHIM) in the temperature range 
$T\approx 10^5-10^6$ K, a gas phase that is expected to contain
a large fraction of the baryons at low redshift.
In this paper we use a hydrodynamical simulation to study 
frequency, distribution, physical conditions, and baryon content 
of the BLAs at $z\approx 0$.
From our simulated spectra we derive a number of BLAs per unit 
redshift of $(d{\cal N}/dz)_{\rm BLA}\approx 38$ for H\,{\sc i} absorbers with 
log $(N$(cm$^{-2})/b$(km\,s$^{-1}))\geq 10.7$, 
$b\geq40$ km\,s$^{-1}$, and total hydrogen column densities 
$N$(H\,{\sc ii}$)\leq 10^{20.5}$ cm$^{-2}$. 
The baryon content of these
systems is $\Omega_{b}$(BLA$)=0.0121\,h_{65}\,^{-1}$, which
represents $\sim 25$ percent of the total baryon budget
in our simulation. 
Our results thus support the idea that BLAs represent
a significant baryon reservoir at low redshift. 
BLAs predominantly trace shock-heated collisionally ionized WHIM gas
at temperatures log $T\approx 4.4-6.2$. About 27 percent of the BLAs in our
simulation originate in the photoionized Ly\,$\alpha$ forest 
(log $T<4.3$) and their large line widths are determined by non-thermal 
broadening effects such as unresolved velocity structure and
macroscopic turbulence.   
Our simulation implies that for a large-enough sample 
of BLAs in FUV spectra it is possible to obtain a 
reasonable approximation of the baryon content 
of these systems solely from the measured H\,{\sc i} column densities 
and $b$ values. 


}

\titlerunning{Simulations of BLAs at low redshift}

\maketitle
%

\section{Introduction}

\begin{figure*}[t!]
\resizebox{1.0\hsize}{!}{\includegraphics{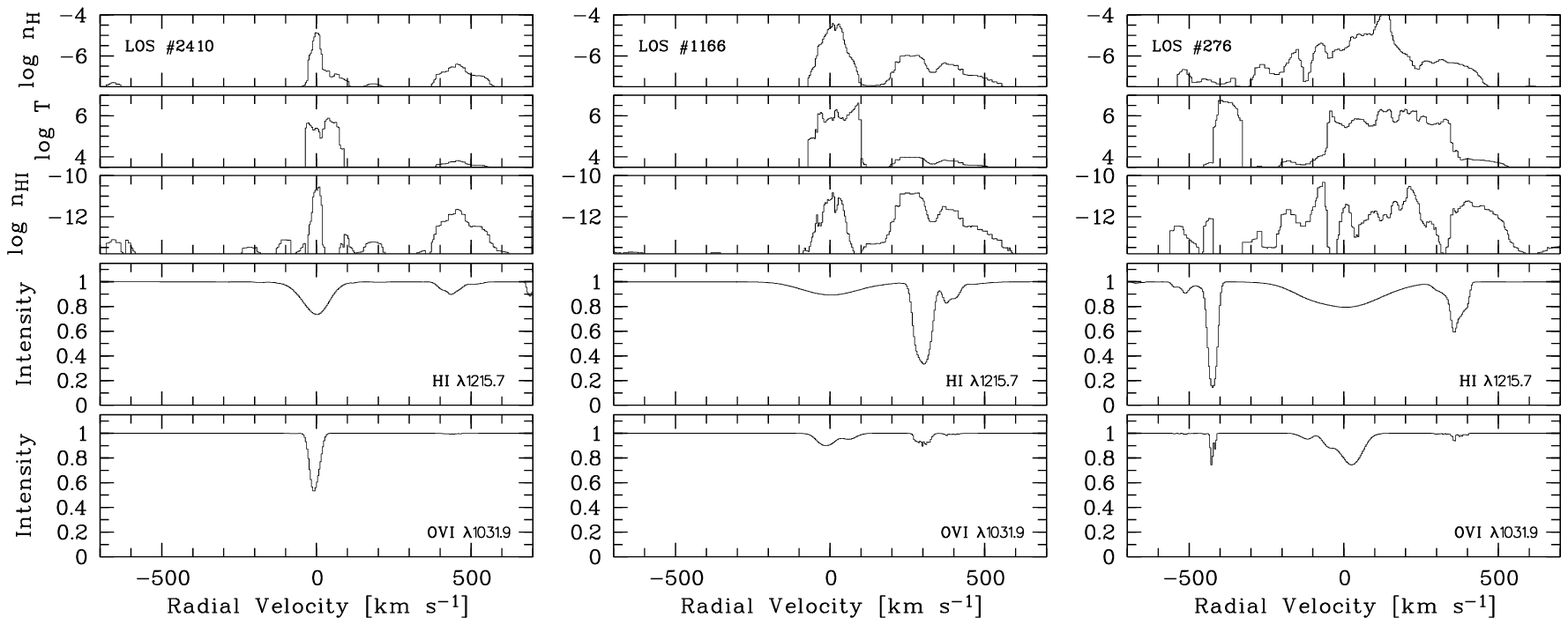}}
\caption[]{Three examples for BLA absorbers in our simulated spectra
are shown. The panels show the logarithmic total hydrogen volume density,
gas temperature, neutral hydrogen volume density, and normalized intensity for
H\,{\sc i} Ly\,$\alpha$ and O\,{\sc vi} $\lambda 1031.9$ absorption
as a function of the radial
velocity along each sightline. 
Densities $n$ and temperatures $T$ are in units cm$^{-3}$
and K, respectively.}
\end{figure*}

Cosmological simulations predict that a large
fraction ($\sim 30-40$ percent) of the baryons
in the local Universe exist in the form of hot
($T=10^5-10^7$ K), highly
ionized intergalactic gas (e.g., Cen \& Ostriker 1999;
Dav\a'e et al.\,2001), 
commonly referred to as the
warm-hot intergalactic medium (WHIM). 
The WHIM is believed to emerge from intergalactic gas
that is shock-heated to high temperatures as the medium is
collapsing under the action of gravity (Valageas, Schaeffer, \& Silk 2002).
With current instrumentation, the most promising 
method to study the distribution and
baryon content of the WHIM is to search for absorption
lines from neutral hydrogen and highly-ionized heavy elements
in far-ultraviolet and X-ray spectra of bright extragalactic 
background sources. 
Space-based observatories such as the {\it Far Ultraviolet
Spectroscopic Explorer} (FUSE), the {\it Hubble Space Telescope}
with its {\it Space-Telescope Imaging Spectrograph} (STIS),
and the {\it Chandra} and {\it XMM-Newton} X-ray observatories have
been used to the study frequency and distribution of highly ionized
oxygen (O\,{\sc vi}, O\,{\sc vii}, O\,{\sc viii}) in the WHIM toward a number
of low-redshift background sources (e.g., Tripp, Savage \& Jenkins 2000; 
Savage et al.\,2002; Fang et al.\,2002; Shull \& Danforth 2005; 
Nicastro et al.\,2005). FUV observations imply that the 
baryon content of O\,{\sc vi} absorbers (which are believed to trace the WHIM
at temperatures around $T=3\times10^5$ K) is
$\Omega_{b}$(O\,{\sc vi}$)\geq 0.0022\,h_{70}\,^{-1}$
(Danforth \& Shull 2005; Sembach et al.\,2004), but this estimate 
depends critically on the assumed oxygen abundance,
which is not well constrained.

Apart from highly ionized oxygen, recent STIS observations 
by Richter et al.\,(2004, 2006) and
Sembach et al.\,(2004) suggest that WHIM filaments 
can be traced by thermally broadened H\,{\sc i}
Ly\,$\alpha$ absorption (see also 
Williger et al.\,2006). Although the vast majority 
of the hydrogen in the WHIM is ionized due to 
particle collisions and the ambient UV background
radiation, a very small fraction ($<10^{-5}$, typically) 
of the hydrogen is expected to be neutral if the gas is 
in ionization equilibrium. For WHIM filaments
with total gas column densities $>10^{19}$ cm$^{-2}$  
the neutral hydrogen should give rise to weak Ly\,$\alpha$
absorption with H\,{\sc i} column densities on the order 
of log $N\approx13$. Due to the high temperature
of the gas ($T=10^5-10^7$ K) and the effect of thermal line broadening,
the Ly\,$\alpha$ absorption is
expected to be very broad and shallow with
large Doppler parameters ($b\geq40$ km\,s$^{-1}$).
About fifty of these broad Ly\,$\alpha$ absorbers (BLAs)
have been identified in STIS low-z QSO spectra so far
(Richter et al.\,2006),
indicating that these systems represent a significant
baryon reservoir at low redshift. The STIS
measurements imply a number of BLAs per unit 
redshift of $(d{\cal N}/dz)_{\rm BLA}\geq 22$ 
for absorbers with log $(N$(cm$^{-2})/b$(km\,s$^{-1}))\geq 11.3$
(detection limit for BLAs in STIS data with a signal-to-noise 
ratio of $\sim 18$ per $10$ km\,s$^{-1}$ wide pixel,
hereafter referred to as (S/N)$_{10}$), and a baryon mass density of 
$\Omega_{b}$(BLA$)\geq 0.0027\,h_{70}\,^{-1}$. The
advantage of using the BLAs for an estimate of $\Omega_{b}$
is that no assumption for the metallicity of the
WHIM gas has to be made and therefore also low-metallicity 
environments can be detected.
Large uncertainties for $\Omega_{b}$(BLA) arise, 
however, from the unknown contribution
of non-thermal line broadening effects and the 
uncertain ionization fraction of the gas (see Richter et al.\,2006).

In this paper we use artificial spectra generated from
a cosmological simulation to study the frequency and
baryon content of BLAs at low redshift. The aim of
this paper is to test whether these simulations 
reproduce the observed spectral signatures of WHIM filaments
(i.e., broad and shallow H\,{\sc i} Ly\,$\alpha$ absorption)
and to investigate the physical conditions in these systems.
This paper is organized as follows: in Sect.\,2 we
explain the simulation method; in Sect.\,3 we describe
the spectral analysis of the artificial spectra; physical
properties of the BLAs are discussed in Sect.\,4; in
Sect.\,5 we estimate the baryon content of the BLAs in
our simulation and compare the results with recent observations;
in Sect.\,6 we compare the relationship between BLAs and
O\,{\sc vi} absorbers; finally, we summarize our study in 
Sect.\,7.      

\section{Simulation method}

For our study we use the output of a numerical simulation
that was part of an investigation of intervening O\,{\sc vi} absorption
arising in WHIM filaments (Fang \& Bryan 2001), and which
is based on a grid-based adaptive mesh refinement method
(Bryan 1996; Norman \& Bryan 1999). The generated cube
is $20\,h^{-1}$ Mpc on a side and has a 
gas and dark matter mass resolution of $6\times10^7$
and $5\times 10^8\,M_{\sun}$, respectively. The 
maximum resolution is $9.8$ kpc (smallest
grid cell). The initial density field was drawn
from an adiabatic cold dark matter power spectrum
as approximated by the relation given in 
Eisenstein \& Hu (1998). 
The simulation is based on a 
cosmological model with $\Omega_{\rm m}=0.4$,
$\Omega_{\rm b}=0.0473$, $\Omega_{\Lambda}=0.6$, and
$h=0.65$ (in units of $100$ km\,s$^{-1}$\,Mpc$^{-1}$).
The inital redshift is $z=50$ and the simulation
stops at $z=0.2$. Radiative heating 
and cooling of the gas is not included
in the simulation, so that the gas temperature in
the unshocked, low-density regions is significantly
underestimated. For these
regions we therefore approximate the 
temperature of the gas using the temperature-density
relation $T=T_0\,(1+\delta)^{\gamma -1}$ found for the 
Ly\,$\alpha$ forest, where $T_0\approx 5000$ K, $\gamma
\approx 1.4$ (e.g., Zhang et al.\,1997) and $\delta$ is
the overdensity of the gas. Note that previous studies 
indicate that radiative heating/cooling
is not expected to significantly influence the evolution
of the WHIM gas at low redshifts (see, e.g., Dav\a'e et al.\,2001).
However, we are planning to investigate this issue
in more detail in a future paper.
A density-dependent metallicity model
was adopted, based on the results from Cen \& Ostriker (1999).
More detailed information about the simulations can be
found in Fang \& Bryan (2001) and Norman \& Bryan (1999)
and references therein. 
The ionization state of the gas in the simulation 
has been modelled using the code CLOUDY 
(version 90.04; Ferland et al.\,1998), which includes
both collisional ionization and photoionization.  
A grid of temperatures
(log $T=3-7$) and densities (log $n_{\rm H}=-8$ to $-2$)
is calculated from CLOUDY assuming an ionizing background
spectrum at $z=0.2$ based on the model of Haardt \& Madau (1996)
together with a mean specific intensity at the Lyman limit
of $J_{\nu}=2 \times 10^{-23}\,$erg\,s$^{-1}$\,Hz$^{-1}$\,sr$^{-1}$.

For the analysis of the simulated region we have
drawn 3000 random lines of sight (LOS) through the
simulated box and have created artificial spectra
for H\,{\sc i} and O\,{\sc vi} following the 
method described in Zhang et al.\,(1997). 
The total redshift path of these 3000 sightlines
is $\Delta z = 20.0$.
The spectra then have been fitted using the 
automatted line fitting algorithm AutoVP
(Dav\a'e et al.\,1997), which delivers
column densities, $N$, and Doppler parameters, $b$,
for the fitted spectral features. 

\section{H\,{\sc i} Spectral analysis}

\subsection{BLA sample selection}

\begin{figure}[t!]
\resizebox{1.0\hsize}{!}{\includegraphics{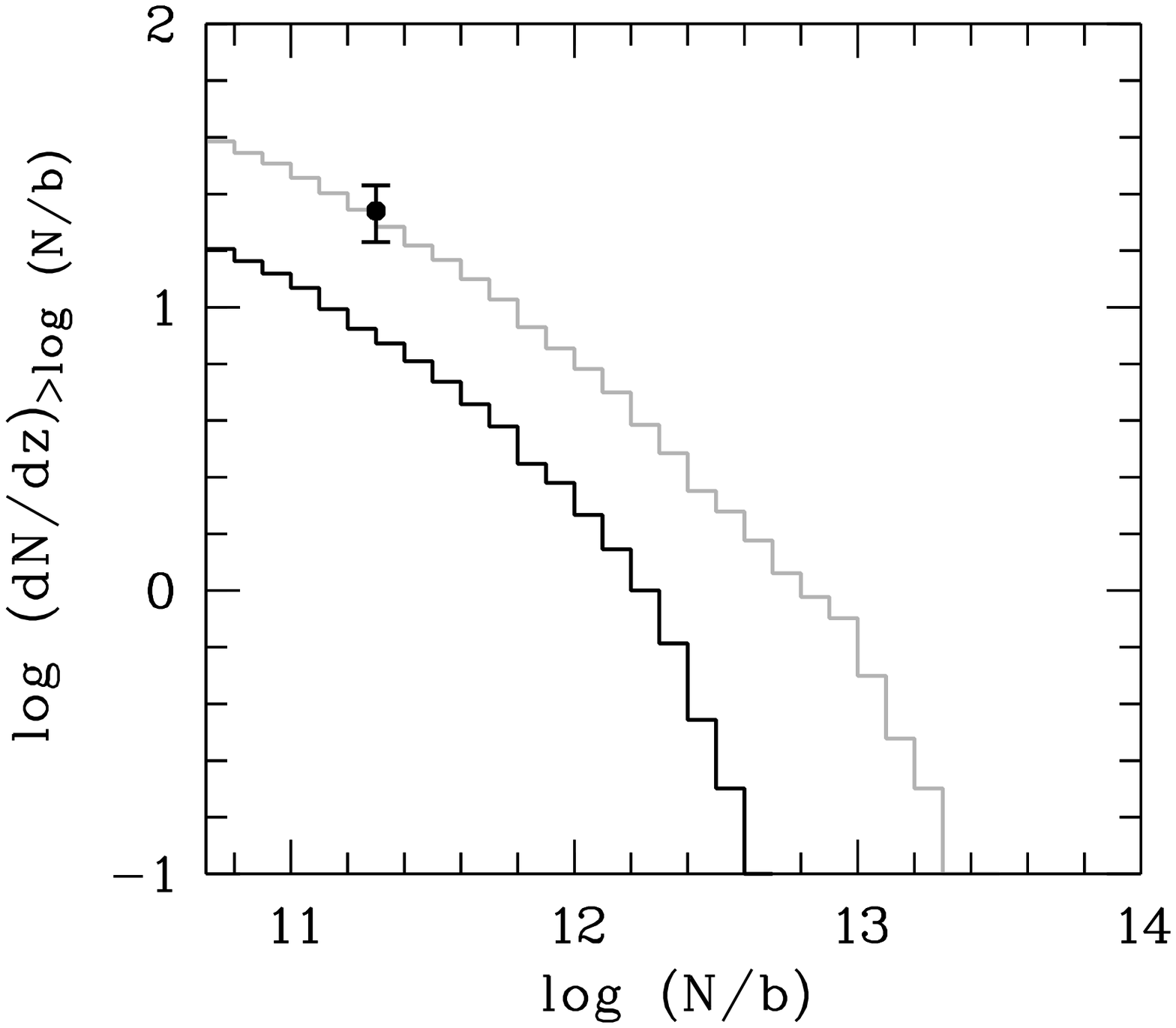}}
\caption[]
{Cumulative number of BLAs per unit redshift, log $(d{\cal N}/dz)_{\rm BLA}$,
as a function of the absorption strength, log $(N$(cm$^{-2})/b$(km\,s$^{-1}))$.
The total sample is shown in gray, the high-quality sample 
is plotted in black. The single data point indicates the
value for $(d{\cal N}/dz)_{\rm BLA}$ from FUV absorption line
measurements with STIS (Richter et al.\,2006).}
\end{figure}

The fitting of the 3000 LOS
with AutoVP results in the detection of 31,340 Ly\,$\alpha$
absorbers. The vast majority of these lines belong to the
photoionized Ly\,$\alpha$ forest. 2671 of these lines have
$b\geq40$ km\,s$^{-1}$ and thus represent BLA candidates
that may trace collisionally ionized gas at temperatures
$T>10^5$ K arising in WHIM filaments. 
We have applied a number of selection criteria to identify
the best BLAs candidates in our sample.  
First, we have excluded broad features that are part of complex
multi-component absorbers, as the determination of line
centers, $b$ values, and column densities for these systems
is afflicted with large systematic uncertainties.
To exclude these crowded regions,
we take into account only those BLA candidate lines whose nearest
neighbouring absorption line is at least $2\sigma$ away, where
$\sigma$ is the standard deviation of a Gaussian-shaped
line defined as:

\begin{equation}
\sigma = \frac{\rm FWHM}{2.354} = 0.705\,b.
\end{equation}

In addition, we have excluded very weak BLA lines by
constraining our BLA sample to H\,{\sc i} absorbers that have
an absorption strength of log $(N$(cm$^{-2})/b$(km\,s$^{-1}))
\geq 10.7$. This limit corresponds to a minimum absorption
depth of $\sim 10$ percent. Note that lines with
log $(N/b)<10.7$ do not significantly
contribute to the baryon content of the BLAs; moreover,
they would clearly fall below the detection limit in
FUV spectra that have signal-to-noise ratios (S/N)$_{10}\leq 50$.
After applying the above selection criteria we
are left with a sample of 770 BLA candidate lines.

BLAs arising in the WHIM probably do not represent single-component, isothermal gaseous
structures, but rather are expected to be multi-phase systems with density substructure
and internally varying ionization conditions. The method of using the BLAs to measure
the baryon content of the WHIM is based on the idea that the shock-heated regions
contain the dominant fraction of the baryons in these
systems, and that these regions also determine the Ly\,$\alpha$ line width due 
to their high temperature and the resulting thermal 
line broadening. In an ideal case, one thus would
expect to observe a single-component, Gaussian-shaped broad absorption feature with
a line-width that serves directly as a measure of the temperature of the gas (see also
Sect.\,3.4). However, if substantial substructure 
is present in a BLA, the thermal line-broadening
is competing with the broadening of the absorption due to unresolved radial-velocity
components. To distinguish between thermal broadening and non-thermal broadening
it is therefore important to look for evidence of velocity-component 
structure in the BLA candidate lines (see also Richter et al.\,2006).
We have done a visual inspection of our 770 BLA candidates to
evaluate the quality of the line fitting (as done by
AutoVP) and to look for evidence for
sub-component structure. 321 BLAs in our
sample are almost perfectly Gaussian shaped and thus
represent excellent candidates for 
thermally broadened WHIM absorbers.
The remaining 449 BLAs show evidence for 
sub-component structure that has been ignored
by the automated line fitting procedure. It could be
that a large fraction
of these broad systems may not be related to the
WHIM but rather represent photoionized multi-component
Ly\,$\alpha$ forest lines.
Note that in real FUV spectra with S/N ratios (S/N)$_{10}<50$ these
multi-component systems would not be distinguishable from the
single-component cases. To compare our results with observational
data it is therefore important to 
assess the properties of {\it all} BLA candidate lines in
our sample.
For the following analysis we distinguish between
the high-quality BLA sample (321 absorbers) and
the total BLA sample (770 absorbers). 

Examples
for BLAs along different lines-of-sight (LOS)
in our simulated spectra are 
presented in Fig.\,1, where we have plotted the total hydrogen
volume density, the gas temperature, the neutral
hydrogen volume density, the normalized H\,{\sc i} 
Ly\,$\alpha$ $\lambda 1215.7$ intensity, and the 
normalized intensity of the O\,{\sc vi} $\lambda 1031.9$
line as a function of the restframe velocity for
three BLAs from the high-quality sample. As 
clearly visible, the H\,{\sc i} Ly\,$\alpha$ absorption
in these systems is broad and symmetric. The left
example (LOS 2410) shows a WHIM BLA with $b\approx 59$ km\,s$^{-1}$
and log $N$(H\,{\sc i}$)=13.37$; the density and 
temperature distribution is very simple in this system.
The width of this BLA line is governed mainly by thermal
broadening of the WHIM gas that has log $T\approx 5.3$. 
The middle panel (LOS 1166) shows a thermally-broadened 
BLA with $b\approx 129$ km\,s$^{-1}$
and log $N$(H\,{\sc i}$)=13.28$. Also this system has a 
simple single-component structure, but the WHIM temperature
is with log $T\approx 5.9$ somewhat higher than in the left 
example, thus resulting in an H\,{\sc i} Ly\,$\alpha$ 
line that is broader (and shallower). The right panel
(LOS 276) shows an extremely broad BLA system with
$b\approx 165$ km\,s$^{-1}$ and log $N$(H\,{\sc i}$)=13.70$. 
Here, the BLA system is characterized by a complex 
density distribution that spans several hundred km\,s$^{-1}$
(see upper panel). The BLA has a high temperature of 
log $T\approx 6.0$, but in this case, the extremely
large width of the line is not dominated by thermal
broadening but by the complex velocity structure.
The shape of the H\,{\sc i} Ly\,$\alpha$ absorption slightly
deviates from a perfect Gaussian, indicating the presence 
of velocity sub-components in this line. Such small 
asymmetries would remain unnoticed in real spectra where
the S/N is limited. 

\subsection{BLA number density}

Fig.\,2 shows the cumulative number of BLAs per unit
redshift, log ($d{\cal N}/dz$), plotted against
the BLA absorption strength, log $(N/b)$.
The total sample (770 systems) is shown in gray,
the high-quality sample (321) is plotted in black.
For $b\geq 40$ km\,s$^{-1}$ and log $(N/b)\geq10.7$
our simulation predicts that $(d{\cal N}/dz)_{\rm BLA}\approx 39$
(total sample) and $(d{\cal N}/dz)_{\rm BLA}\approx 16$ (high-quality
sample). For log $(N/b)\geq11.3$ the BLA number density
in the total sample reduces to $(d{\cal N}/dz)_{\rm BLA}\approx 19$.
This is very similar to the value obtained by recent
STIS observations of BLAs towards low-redshift
quasars ($(d{\cal N}/dz)_{\rm BLA}=22$ for reliably detected
systems; Richter et al.\,2006). The value for $(d{\cal N}/dz)_{\rm BLA}$ from
the STIS measurements is indicated in Fig.\,2.

\subsection{H\,{\sc i} column densities}

\begin{figure}[t!]
\resizebox{1.0\hsize}{!}{\includegraphics{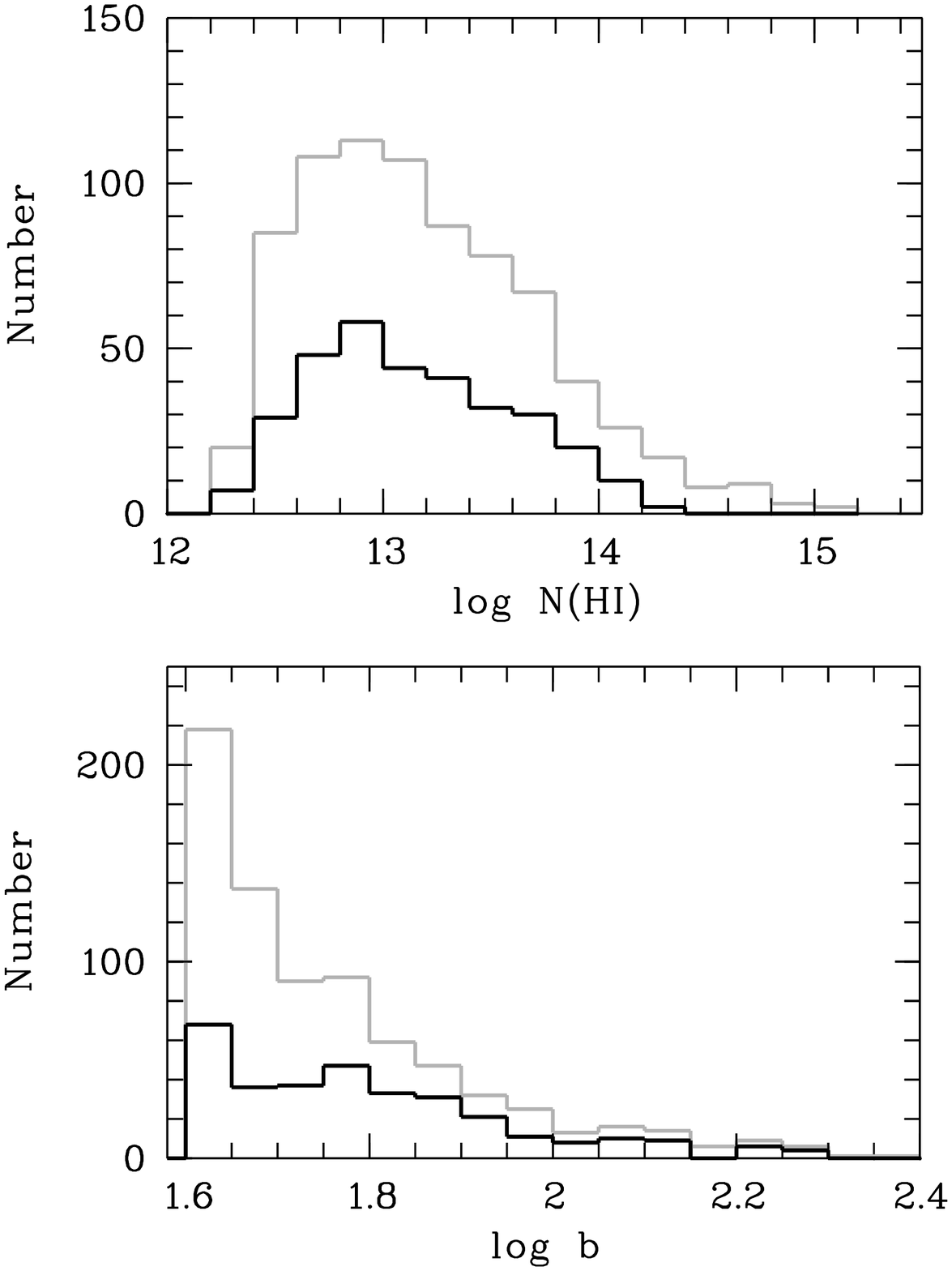}}
\caption[]
{H\,{\sc i} column density and $b$ value distribution resulting
from the Voigt fit of the BLA lines in the simulation. The
gray line shows the distribution of the total sample (770 systems),
the black line shows the distribution for the restricted sample
(321 systems). Column densities are in units cm$^{-2}$, $b$ values
are in units km\,s$^{-1}$.}
\end{figure}

Fig.\,3 shows the distribution of logarithmic H\,{\sc i}
column densities and logarithmic H\,{\sc i} $b$ values,
as obtained from the
profile fitting of the BLA candidate lines.
The total sample is shown in gray, the high-quality sample
is shown in black.
The H\,{\sc i} column density distribution (Fig.\,3, upper panel)
shows that the majority of the BLA systems ($\sim 92$ percent)
have relatively low H\,{\sc i} column densities with
log $N$(H\,{\sc i}$)\leq 14$ and a peak near
log $N$(H\,{\sc i}$)\approx 13$. Note that for column densities
lower than that our sample is incomplete due to the
exclusion of systems with log $(N/b)<10.7$ (see Sect.\,3.1).
The true H\,{\sc i} column density distribution of the BLAs
most likely further rises towards lower column densities.
The distribution of BLA H\,{\sc i} column densities 
in our simulation is qualitatively very similar to the one observed in
STIS spectra of low$-z$ quasars (Richter et al.\,2006; their Fig.\,5).

\subsection{H\,{\sc i} $b$ values}

The BLA $b$ value distribution is plotted in the lower panel
of Fig.\,3. Some BLAs in our simulation have $b$ values larger
than $200$ km\,s$^{-1}$, but the majority of the systems
have relatively moderate line widths with $b$ values ranging
from $40$ to $100$ km\,s$^{-1}$ (log $b=1.6-2.0$).
The $b$ values of the BLAs are assumed to be composed
of a thermal component, $b_{\rm th}$, and a non-thermal
component, $b_{\rm non-th}$, in the way that

\begin{equation}
b=\sqrt{b_{\rm th}\,^2+b_{\rm non-th}\,^2}.
\end{equation}

The non-thermal component may include processes like
macroscopic turbulence, unresolved velocity-components,
and others (see Richter et al.\,2006 for a detailed discussion). 
The contribution of the thermal component to $b$
depends on the gas temperature:

\begin{equation}
b_{\rm th} =  \sqrt \frac{2kT}{m} \approx 0.129 \sqrt \frac{T}{A}\, \rm{km\,s}^{-1},
\end{equation}

where $k$ is the Boltzmann constant, $m$ is the particle mass,
and $A$ is the atomic weight.
For the shock-heated WHIM gas with log $T\geq5$ one thus expects $b_{\rm th}\geq40$
km\,s$^{-1}$. The non-thermal broadening mechanisms
are expected to contribute to some degree to the total 
$b$ values in WHIM absorbers (see Richter et al.\,2006), so that
the measured $b$ value of a BLA provides only an upper limit for the 
temperature of the gas.
If we compare the $b$ value distribution
of the total BLA sample in our simulation
(lower panel in Fig.\,2, gray line) with
that of the high-quality sample (black line), we see
that the fraction of absorbers with log $b<1.8$ ($b<63$
km\,s$^{-1}$) is very large in the total sample ($\sim 70$ percent),
while for the high-quality sample it is somewhat 
smaller ($\sim 59$ percent).
This implies that 
for low $b$ values the contamination from multi-component
Ly\,$\alpha$ absorbers (i.e., systems for which $b_{\rm non-th}$
is dominant) in the total BLA sample is substantial
(see also Sects.\,3.1, 4.1, and 5.1).

\section{Physical properties of BLAs}

\subsection{Temperature distribution}

\begin{figure}[t!]
\resizebox{1.0\hsize}{!}{\includegraphics{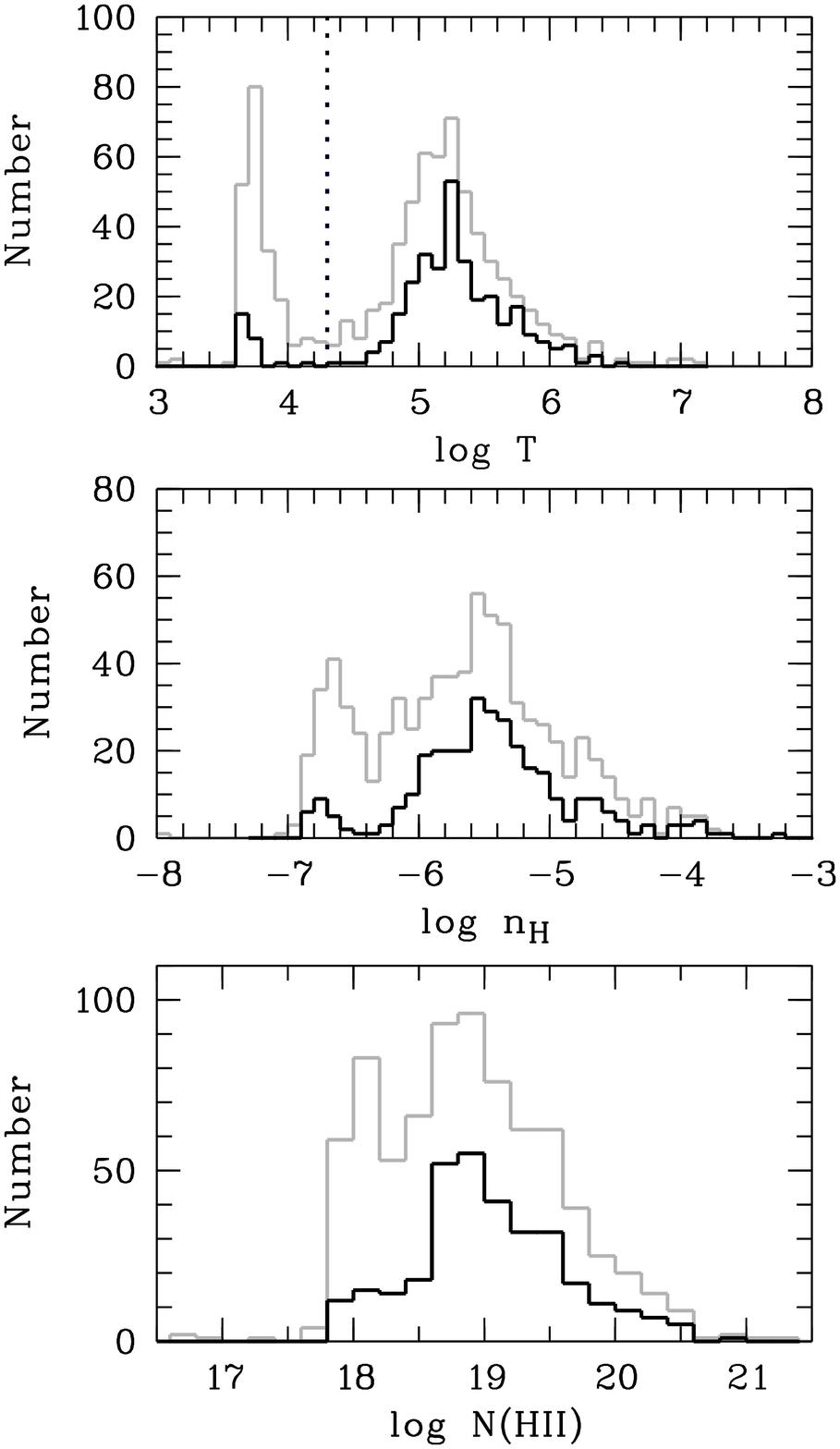}}
\caption[]
{Distribution of temperatures ($T$ in K), hydrogen volume densities
($n$ in cm$^{-3}$), and
total hydrogen column densities ($N$ in cm$^{-2}$) 
for the total BLA sample (770
absorbers; shown in gray) and the high-quality sample (321 absorbers;
shown in black).}
\end{figure}

We have determined for each BLA in our sample a characteristic
temperature by calculating the 
mean of the temperatures in the cells that contribute
to the H\,{\sc i} Ly\,$\alpha$ absorption,
weighted by H\,{\sc i} optical depth in each cell.
The H\,{\sc i} weighting is a crucial point, since 
it is the temperature in the cells with the largest
H\,{\sc i} optical depths that determines the thermal
broadening (and thus the width) of the Ly\,$\alpha$ absorption.
In Fig.\,4, upper panel, we show the distribution
of log $T$ for the total sample (gray)
and the high-quality sample (black). The BLA temperature
distribution is bimodal, with most BLAs spanning a range of
$4.4 \leq {\rm log}\,T \leq 6.2$ with a peak near log $T\approx5.2$,
thus typical for what is expected for the low-temperature
WHIM (e.g., Dav\a'e et al.\,2001).
A second sharp peak is visible at lower temperatures near
log $T=3.8$, at temperature that is characteristic
for the photoionized Ly\,$\alpha$ forest (e.g., Dav\a'e \& Tripp 2001).
At such low temperatures, thermal broadening
contributes only with $b_{\rm th}\approx 10$ km\,s$^{-1}$ to
the total line width of these systems 
(see equation (3)). The measured large
width of these lines ($b\geq 40$ km\,s$^{-1}$) therefore must
be determined by non-thermal processes, such as unresolved
component structure or macroscopic turbulence.

The BLA temperature distribution clearly separates
low-temperature, photoionized H\,{\sc i} absorbers with
log $T<4.3$ (i.e., broad 
Ly\,$\alpha$ forest lines) from shock-heated,
high-temperature WHIM absorbers that have log $T\geq4.3$
(indicated by the dotted line in Fig.\,4, upper panel).  
While for the high-quality
sample only about eight percent of the BLAs have 
temperatures log $T<4.3$, the fraction of BLAs 
with log $T<4.3$ in the total sample is $\sim 27$ percent.
Nearly all (198 out of 209) of these low-temperature BLAs in the 
total sample have $b$ values $\leq 63$ km\,s$^{-1}$ 
(log $b\leq1.8$).
This shows that BLA absorbers
with $b$ values $<63$ km\,s$^{-1}$ in the total BLA
sample are substantially contaminated by photoionized
Ly\,$\alpha$ forest lines (see also Fig.\,3 and Sect.\,3.1), while
for BLAs with $b\geq 63$ km\,s$^{-1}$ the collisionally
ionized WHIM absorbers dominate the BLA population.

Note that due to the relatively small box size and the
missing large-scale power in our simulation the characteristic temperatures
of WHIM filaments and BLAs may be slightly underestimated. 
We are planning to examine this issue in more detail
in a future paper using a higher-resolution simulation.

\subsection{Density distribution}

The distribution of logarithmic hydrogen volume densities
($n_{\rm H}$ in units cm$^{-3}$) in our two BLA samples
is shown in Fig.\,4, middle panel. 
In a similar way as for the temperatures, we have 
calculated for each BLA $n_{\rm H}$ from the mean of the
hydrogen volume densities in the cells that contribute
to the H\,{\sc i} absorption,
weighted by the H\,{\sc i} optical depth in each cell.
The distribution of log $n_{\rm H}$ spans a large range
between $-3.6$ and $-7.2$. 
Similar to the temperature distribution, the
distribution of log $n_{\rm H}$ is bimodal with one
peak near log $n_{\rm H}\approx-5.4$ and a second
one near log $n_{\rm H}\approx-6.7$. Again, this
bimodal distribution separates the low-density, photoionized
Ly\,$\alpha$ forest lines (log $n_{\rm H}<-6.2$, typically) from 
the higher-density WHIM absorbers (log $n_{\rm H}\geq-6.2$, typically).
The above density peaks near log $n_{\rm H}=-6.7$ and
$-5.4$ correspond to overdensities of
$\rho_{\rm H}/\overline{\rho_{\rm H}} 
\sim 1$ for the low-density Ly\,$\alpha$ forest
and $\rho_{\rm H}/\overline{\rho_{\rm H}} \sim 20$ 
for the WHIM absorbers, thus
similar to what has been found in other simulations
(e.g., Dav\a'e \& Tripp 2001; Dav\a'e et al.\,2001).

\subsection{Ionized hydrogen column densities}

The column density of ionized hydrogen in each BLA
($N$(H\,{\sc ii}$)\approx N$(H))
can be estimated directly
from our simulation,
i.e., by integrating the
hydrogen volume density, $n_{\rm H}$, over the
cells that are located within the $2\sigma$ 
velocity range of each BLA (see equation (1)).
In the lower panel of Fig.\,4 we show the distribution
of log $N$(H\,{\sc ii}) for the total sample (gray)
and the high-quality sample (black). Most BLAs have
ionized (=total) hydrogen column densities in
the range log $N$(H\,{\sc ii}$)=18-20$ with a peak
near log $N$(H\,{\sc ii}$)=18.8$. Less than seven percent 
of the BLAs have
H\,{\sc ii} column densites larger than $10^{20}$ cm$^{-2}$,
but note that these systems contain a substantial
fraction of the total gas column density in the BLAs
($\sim 57$ percent in the total sample, $\sim 51$ percent
in the high-quality sample). 87 percent of the broad Ly\,$\alpha$
forest lines with log $T<4.3$ have H\,{\sc ii} column densities
log $N$(H\,{\sc ii}$)<18.5$. Broad absorbers that arise
in the WHIM therefore dominate the distribution of the BLA H\,{\sc ii}
column densities in the range log $N$(H\,{\sc ii}$)=18.5-21.0$.

\subsection{Ionization fraction}

\begin{figure}[t!]
\resizebox{1.0\hsize}{!}{\includegraphics{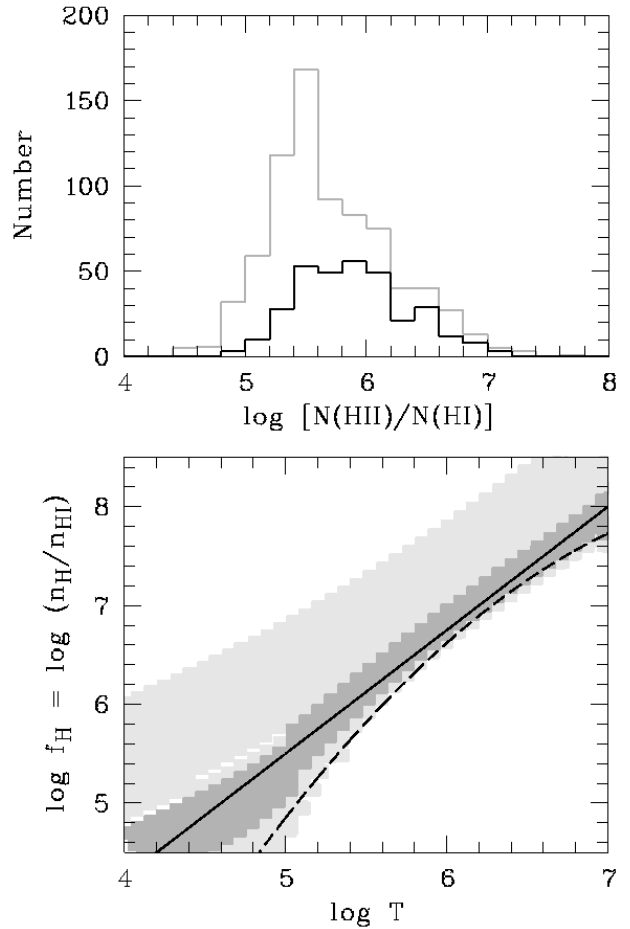}}
\caption[]
{{\it Upper panel:} distribution of ionization fractions
in the total BLA sample (gray) and the high-quality sample (black).
{\it Lower panel:} the hydrogen ionization fraction for each cell 
in our simulation (photoionization+collisional ionization),
log $f_{\rm H}={\rm log} (n_{\rm H}/n_{\rm HI})$, 
is plotted as a function of the gas temperature, log ($T$(K)). 
The light gray shaded area indicates cells in the density range
log $n_{\rm H}=-5$ to $-7$. The dark gray shaded area refers to
cells that have log $n_{\rm H}=-5.3$ to $-5.6$, thus a density
range that is characteristic for BLAs. The solid black line shows the
best linear fit of the data in the BLA density range, the 
dashed black curve shows a function that approximates
log $f(T)$ for a pure collisional ionization model.}
\end{figure}

In the upper panel of Fig.\,5 we have plotted the distribution
of the ionized hydrogen ionization
fractions, $N$(H\,{\sc ii}$)/N$(H\,{\sc i}), for our two
BLA samples. The logarithmic ionization fractions in both samples
typically range between 5 and 7 with a peak near $5.6$. 

In the following we want to explore in more detail the 
individual ionization mechanisms (collisional ionization and 
photoionization) and their role in the observed
$N$(H\,{\sc ii}$)/N$(H\,{\sc i}) distribution.
For pure collisional ionization and in ionization 
equilibrium, the hydrogen ionization fraction,
log $f_{\rm H}={\rm log}\,(n_{\rm H}/n_{\rm HI})$, is governed solely
by the gas temperature (e.g., Sutherland \& Dopita 1993). The ionization
fraction then can be approximated via

\begin{equation}
{\rm log}\,f_{\rm H}(T)\approx -13.9 + 5.4\,{\rm log}\,T - 0.33
({\rm log}\,T)^2.
\end{equation}

However, as the densities in the WHIM absorbers 
are low (log $n_{\rm H}<-5$, typically;
see Sect.\,4.2), it is likely that photoionization 
contributes to the 
ionization fraction of the gas.
To assess the temperature dependence
of $f_{\rm H}$ in our combined collisional+photoionization model
we have plotted in Fig.\,5, lower panel, for each of the 
$\sim 1.5\times 10^6$ cells along our 3000 LOS the
ionization fraction, log $f_{\rm H}$, as
a function of the gas temperature, log $T$. The light gray shaded area shows
cells with hydrogen volume densities in the range log $n_{\rm H}=-5$ to $-7$. 
The dark gray shaded area shows cells with log $n_{\rm H}=-5.3$ to $-5.9$,
thus in a density range that is characteristic for the shock-heated BLAs (see 
Fig.\,3). The dashed black curve indicates the $f_{\rm H}-T$ relation for
pure collisional ionzation (equation (4)). As clearly visible, all points
lie above the relation for collisionally ionized gas, demonstrating that
photoionization is important. For the density range that is typical for the BLAs
in our simulation (log $n_{\rm H}=-5.3$ to $-5.9$; dark gray shaded area)
the combined collisional+photoionization model suggests that 
log $f_{\rm H}$ is roughly linearly related to log $T$, so that

\begin{equation}
{\rm log}\,f_{\rm H}(T)\approx -0.75 + 1.25\,{\rm log}\,T.
\end{equation}

This approximation is shown in Fig.\,5, lower panel, as a solid black line.

\subsection{Non-thermal line broadening in BLAs}

\begin{figure}[t!]
\resizebox{1.0\hsize}{!}{\includegraphics{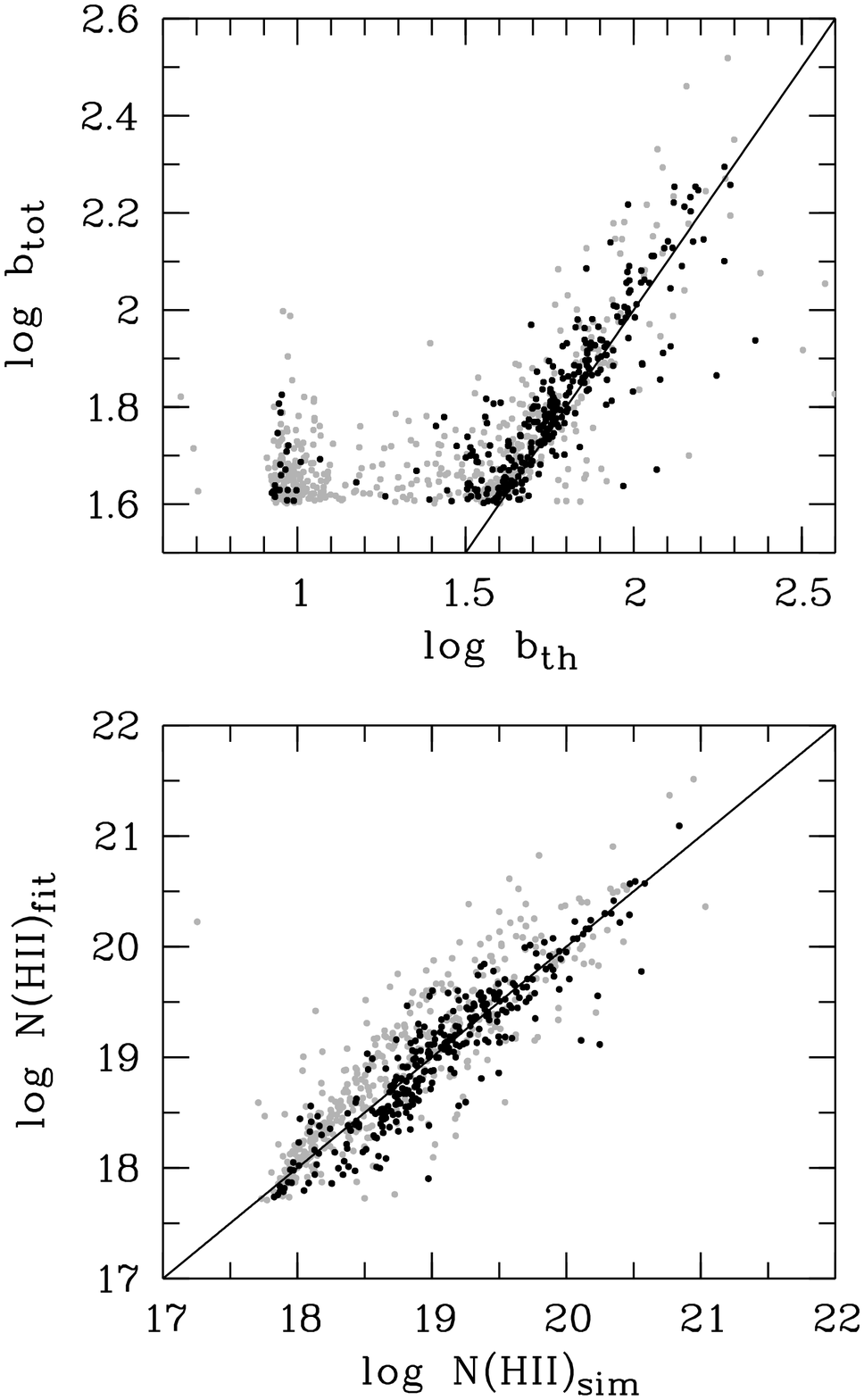}}
\caption[]
{{\it Upper panel:} The total $b$ values {($b$ in units
km\,s$^{-1}$)} for the BLAs as
obtained from the profile fitting is plotted against the
thermal component, $b_{\rm th}$, as calculated for each
BLA from the H\,{\sc i} optical depth weighted temperature
in each system (Sect.\,4.1). The solid black line
indicates $b_{\rm tot}=b_{\rm th}$.
{\it Lower panel:} comparison between the total hydrogen
column densities ($N$ in units cm$^{-2}$) estimated indirectly from the measured
line widths and H\,{\sc i} column densities (log $N$(H\,{\sc ii}$)_{\rm fit}$;
see Sect.\,4.6) and derived directly from the simulation
(log $N$(H\,{\sc ii}$)_{\rm sim}$; Sect.\,4.3). The black
solid line indicates log $N$(H\,{\sc ii}$)_{\rm fit}=$\,log
$N$(H\,{\sc ii}$)_{\rm sim}$. 
}
\end{figure}

In Fig.\,6, upper panel, we have plotted the logarithmic $b$ values
from the line fitting of our BLA sample against log $b_{\rm th}$, which 
represents the thermal contribution to $b$ for each BLA, as calculated 
from the BLA temperatures (Sect.\,4.1) and the inversion of equation 
(3). The gray dots show the data points for our total BLA sample, the black
dots show the data points for the high-quality sample.
For pure thermal broadening, $b=b_{\rm th}$ (equation (2)). This 
relation is shown as solid black line in the upper panel of Fig.\,5.
As expected from the BLA temperature distribution (Fig.\,4),
there are a number of low-temperature, photoionized systems
that have $b_{\rm th} \ll b$ because 
of non-thermal line broadening effects; 
these systems cluster near log $b_{\rm th}=1$.
The BLAs that refer to the shock-heated 
WHIM roughly follow the slope of the relation
$b=b_{\rm th}$, but are systematically offset by a 
few percent. A statistical
comparison between $b$ and $b_{\rm th}$ for 
log $b_{\rm th}\geq 1.5$ implies that for both the total
sample and the high-quality sample the relation 
$b=1.1\,b_{\rm th}$
gives the best fit to the data. This means that non-thermal broadening
mechanisms contribute (on average) with $\sim 10$ percent to the
total $b$ value of a WHIM BLA in our simulation. From equation (2)
then follows that (on average) 
$b_{\rm non-th}\approx 0.5\,b_{\rm th}$ for these systems.
Note that there are a number of systems that have
$b<b_{\rm th}$. Although this appears nonsensical
at first, an inspection of these systems suggests that
their absorption profile is dominated by a cooler
gas component that has a relatively large H\,{\sc i}
optical depth. 
The fitted $b$ value in these systems
only reflects the temperature of this cooler gas component, 
but is not good measure for the temperature of the 
hot gas, whose absorption is hidden in a broad wing that is ignored
by AutoVP. Although $b_{\rm th}$ is based on
the optical-depth weigthed temperature of the gas in our BLAs, 
$b<b_{\rm th}$ can occur in a few systems where this 
effect is particularly pronounced.

\subsection{Indirect estimate of the ionized gas content in BLAs}

The combination of the equations (2), (3), (5), 
and $b=1.1\,b_{\rm th}$ allows us to estimate
the ionization fraction, $f_{\rm H}$, in a thermally 
broadened BLA indirectly from its
measured $b$ value. Together with the measured H\,{\sc i} column density
one thus can estimate the total hydrogen column density in a BLA
via $N$(H\,{\sc ii}$)=f_{\rm H}\,N$(H\,{\sc i}). Combining all these
equations gives

\begin{equation}
{\rm log}\,N({\rm H\,II})\approx {\rm log}
\,N({\rm H\,I})+1.25\,{\rm log}\,(0.81b^2)+1.47.
\end{equation}

Since $b$ and 
$N$(H\,{\sc i}) are the only parameters of BLAs that can be measured
in real FUV spectra, this indirect estimate of $N$(H\,{\sc ii}) is 
the only possible method to observationally constrain 
the total gas budget and thus the baryon content of BLAs in the
low-redshift Universe (see Richter et al.\,2006). 
With our simulation we now want to examine
the reliability of this indirect method for our sample of BLA absorbers.
In Fig.\,6, lower panel, we have plotted $N$(H\,{\sc ii}$)_{\rm fit}$ against
$N$(H\,{\sc ii}$)_{\rm sim}$
for the total sample (gray dots) and the high-quality sample (black dots).
$N$(H\,{\sc ii}$)_{\rm fit}$ is the hydrogen column density 
derived indirectly via equation (6) and $N$(H\,{\sc ii}$)_{\rm sim}$ is 
the hydrogen column density obtained directly from the simulation using
the cell integration (Sect.\,4.3). The black solid line indicates the ideal case,
$N$(H\,{\sc ii}$)_{\rm fit}=N$(H\,{\sc ii}$)_{\rm sim}$. Although the
data points show a significant scatter around that line, the plot 
implies that (for a large enough sample) the indirect method indeed
serves as a valuable tool to roughly estimate the total hydrogen
column densities in BLAs. For the total sample the mean scatter is $0.25$ dex,
thus larger than for the restricted sample ($0.18$ dex) owing to the fact
that non-thermal line broadening processes in these systems
cause a considerable uncertainty for the indirect estimate of $N$(H\,{\sc ii}).

\section{The baryon content of BLAs}

\begin{figure*}[t!]
\resizebox{1.0\hsize}{!}{\includegraphics{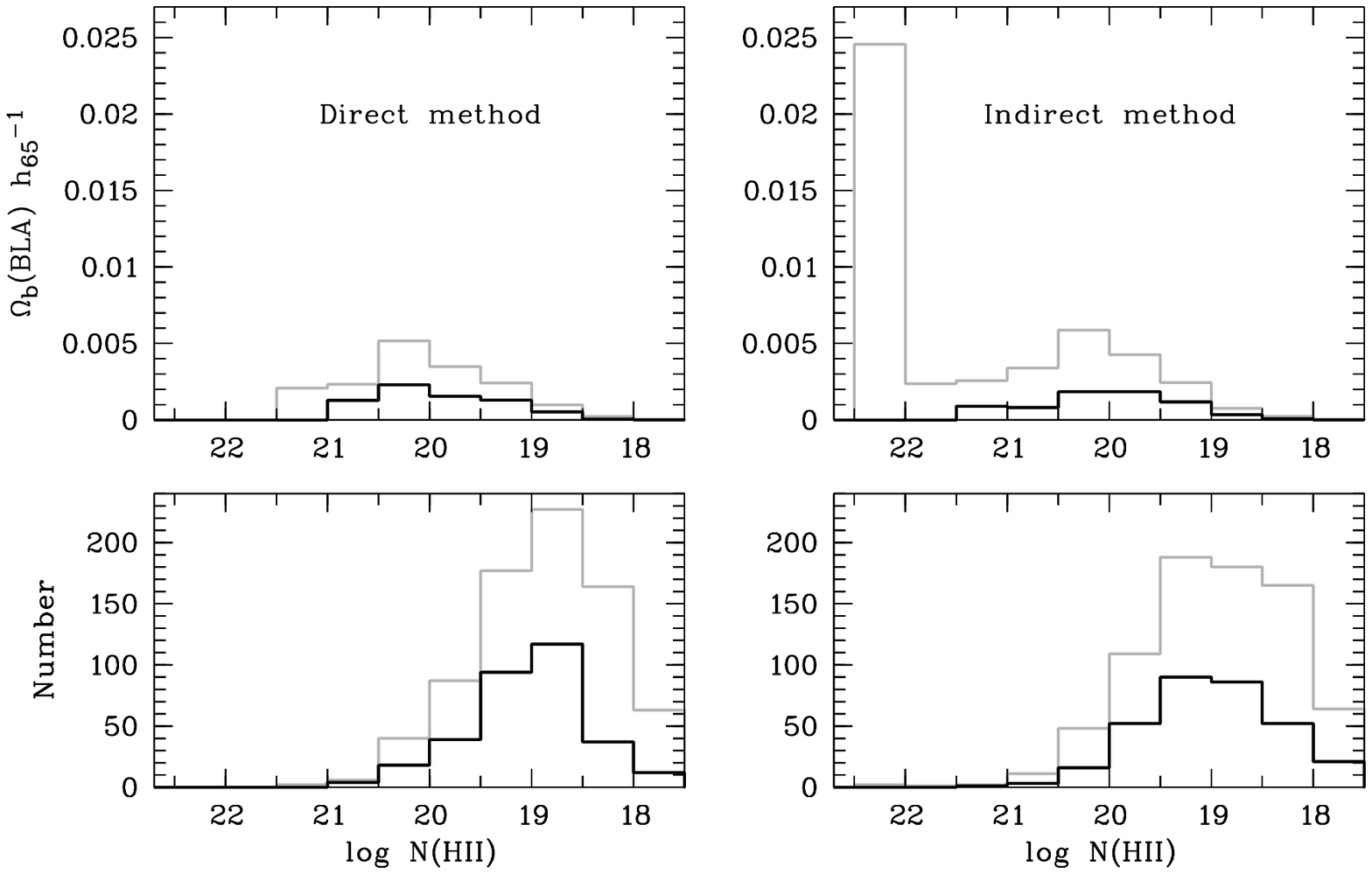}}
\caption[]
{
Distribution of $\Omega_b$(BLA)
and BLA numbers for bins of log $N$(H\,{\sc ii}), where
$N$ is in units cm$^{-2}$. The 
left panel shows the distribution based on the 
direct method to estimate $N$(H\,{\sc ii}) (cell
integration), the right panel shows the distribution
based on the indirect method (equation (6)).
The bin size for log $N$(H\,{\sc ii}) is $0.5$ dex. 
}
\end{figure*}

\subsection{The baryon content of BLAs in the simulation}

From the cell integration (direct method; Sect.\,4.3) 
we derive a total hydrogen column 
density of $2.31 \times 10^{22}$ cm$^{-2}$ for the total BLA sample (770 systems)
and $9.67 \times 10^{21}$ cm$^{-2}$ for the high-quality sample (321 systems).
The cosmological mass density of the BLAs
in terms of the current critical density $\rho_{\rm c}$ is given by
\begin{equation}
\Omega_b{\rm (BLA)}=\frac{\mu\,m_{\rm H}\,H_0}
{\rho_{\rm c}\,c}\,N({\rm HII})_{\rm total}\,\Delta X^{-1},
\end{equation}
where $\mu=1.3$, $m_{\rm H}=1.673 \times 10^{-27}$ kg, $H_0=
65$ km\,s$^{-1}$\,Mpc$^{-1}$, and $\rho_{\rm c}=3H_0\,^2/8 \pi G$.
To compare our results with recent observations we assume 
at this point that the comoving path length for our BLA sample
is $\Delta X=\Delta z=20.0$ (i.e., we assume that $q_0=0$). 
Using equation (7), the values for $N$(H\,{\sc ii}$)_{\rm total}$ 
listed above correspond to baryon densities of 
$\Omega_b$(BLA$)=0.0167\,h_{65}\,^{-1}$ and $0.0070\,h_{65}\,^{-1}$, respectively.
From this follows that the BLAs trace a considerable fraction of the 
baryons in our simulation ($\sim 35$  percent in the total sample, $\sim 15$ 
percent in the high-quality sample). Note that 
the contribution to $\Omega_b$(BLA) from photoionized absorption 
systems with log $T<4.3$ (Ly\,$\alpha$ forest) is negligible ($\sim 2$ percent
in the total sample and $<1$ percent in the high-quality sample).
In the upper left panel of Fig.\,7 we show the distribution of 
baryons as a function of the total hydrogen column density 
for the direct method. The gray
line refers to the total sample, the black line indicates the 
high-quality sample. Values for log $N$(H\,{\sc ii}) are given
in $0.5$ dex-wide bins and the values for $\Omega_b$(BLA) indicate
the baryon content per bin. The lower left panel of 
Fig.\,7 shows the number of BLAs in each column-density bin.
From this comparison we learn that BLA systems with 
log $N$(H\,{\sc ii}$)=19.5-20.5$
contribute most to $\Omega_b$(BLA)
($\sim 56$ percent), while they contribute with only $\sim 16$ percent
to the total number of BLAs. 

\subsection{The baryon content of BLAs derived from the 
indirect estimate}

Using the indirect method for the determination of $N$(H\,{\sc ii})
(equation (6)), we obtain values of 
$\Omega_b$(BLA$)=0.0465\,h_{65}\,^{-1}$ (total sample)
and $0.0070\,h_{65}\,^{-1}$ (high-quality sample). 
Thus, while direct and indirect method give identical
results for the high-quality sample, the indirect method
vastly overestimates $\Omega_b$(BLA) for the total sample. 
In the right panel of Fig.\,7 we plot (in a way similar 
as for the direct method) the distribution of $\Omega_b$(BLA)
and BLA numbers for bins of log $N$(H\,{\sc ii}). When 
compared to the left panel, it immediately becomes clear that the
indirect method wrongly suggests the presence of a few (three) BLA
systems with very large total hydrogen column densities 
(log $N$(H\,{\sc ii}$)>21.5$) that would contain more baryonic
matter than all the other BLAs together.
This is also seen in the lower panel of Fig.\,6, where most of
the few points with log $N$(H\,{\sc ii}$)_{\rm sim}\geq 20.5$
clearly lie above the expected relation 
$N$(H\,{\sc ii}$)_{\rm fit}=N$(H\,{\sc ii}$)_{\rm sim}$. 
A visual inspection of these BLA systems shows that all but one
systems are multi-component absorbers whose widths are determined
mostly by non-thermal line broadening. The velocity-component
structure is not properly accounted for in the fit by AutoVP,
resulting in large discrepancies between  
$N$(H\,{\sc ii}$)_{\rm fit}$ and $N$(H\,{\sc ii}$)_{\rm sim}$
for these systems. As these
high-column density absorbers dominate $\Omega_b$(BLA), the 
overestimate of $N$(H\,{\sc ii}) for these systems 
with the indirect method together with 
the low-number statistics in the high-column density regime
introduces a substantial uncertainty for the (indirect) determination 
of $\Omega_b$(BLA).
The discrepancy in $\Omega_b$(BLA) between the direct and indirect method
becomes less severe if we ignore these few high-column
density systems. If we restrict our total BLA 
sample to values of log $N$(H\,{\sc ii}$)_{\rm fit}
\leq 20.5$ (754 systems), we derive values of $\Omega_b$(BLA$)=0.0121\,h_{65}\,^{-1}$
(cell integration) and $0.0136\,h_{65}\,^{-1}$ (indirect method), thus a much
better agreement between these two methods.
In Table 1 we compare values for $\Omega_b$(BLA) from 
the direct and the indirect method for a number of different 
selection criteria. From this follows that the agreement 
between direct and indirect method
actually is quite good if one ignores the very broadest BLAs that 
have $b$ values larger than $200$ km\,s$^{-1}$.

Following the statistics in our simulation, the best range for
the analysis of BLA systems in future high-resolution, high S/N FUV 
spectra is $b=60-200$ km\,s$^{-1}$. For this range, our analysis
yields $(d{\cal N}/dz)_{\rm BLA}\approx 13$ and values for $\Omega_b$(BLA) of 
$0.0104\,h_{65}\,^{-1}$ (cell integration) and 
$0.0111\,h_{65}\,^{-1}$ (indirect method).
Thus, while the agreement between direct and indirect method
is excellent for these $b$ values, the contamination of the
BLA number density by
photoionized Ly\,$\alpha$ forest lines with log $T<4.3$ is 
negligible ($\sim 5$ percent).

\begin{table*}[t!]
\caption[]{Distribution of the baryons in BLAs with 
log $(N$(cm$^{-2})/b$(km\,s$^{-1}))\geq 10.7$}
\begin{tabular}{lrrrr}
\hline
Selection                  & $N$    & $d{\cal N}/dz$ &
$\Omega_b$(BLA$)_{\rm sim}\,h_{65}$ & $\Omega_b$(BLA$)_{\rm fit}\,h_{65}$ \\
\hline
Total sample               & 770    &   39    &   0.0167  &   0.0465 \\
High-quality sample        & 321    &   16    &   0.0070  &   0.0070 \\
\hline
$b=40-100$ km\,s$^{-1}$    & 700    &   35    &   0.0076  &   0.0102 \\
$b=100-200$ km\,s$^{-1}$   &  64    &    3    &   0.0064  &   0.0068 \\
$b>200$ km\,s$^{-1}$       &   6    & $<1$    &   0.0027  &   0.0295 \\
$b=60-200$ km\,s$^{-1}$    & 268    &   13    &   0.0104  &   0.0111 \\
\hline
log $N$(H\,{\sc ii}$)_{\rm fit} \leq 20.5$    & 754 &  38 &  0.0121 & 0.0136 \\
log $N$(H\,{\sc ii}$)_{\rm fit} \leq 20.0$    & 706 &  35 &  0.0076 & 0.0077 \\
log $N$(H\,{\sc ii}$)_{\rm fit} \leq 19.5$    & 597 &  30 &  0.0038 & 0.0035 \\
log $N$(H\,{\sc ii}$)_{\rm fit} \leq 19.0$    & 409 &  20 &  0.0012 & 0.0010 \\
\hline
\end{tabular}
\end{table*}

\subsection{Comparison with FUV observations}

Richter et al.\,(2006) have studied properties and baryon
content of BLAs in a sample of four low-redshift
quasars using STIS high-resolution spectra. Along a total
redshift path of $\Delta z=0.928$ they detect 49 BLA
candidate systems above a sensitivity limit of
log $(N/b)\approx 11.3$. However, only 20 of these candidates
with $b<100$ km\,s$^{-1}$ are considered as secure detections.
From these 20 BLAs, Richter et al.\,(2006) obtain values
of $(d{\cal N}/dz)_{\rm BLA}=22$ and $\Omega_b$(BLA$)=0.0025\,h_{65}\,^{-1}$.
It is important to note that their estimate of $\Omega_b$(BLA) 
is based on the assumption that the gas is in collisional
ionization equilibrium. Yet, our simulation shows that 
photoionization is important when it comes to the estimate
of total hydrogen column densities and $\Omega_b$(BLA). 
If we take the 20 BLA candidates (secure detections) in the 
STIS data presented
by Richter et al.\,(2006) and transform their measured 
H\,{\sc i} column densities and $b$ values (their Table 4) into total
gas column densities using equation (6) (i.e., now taking
both collisional ionization and photoionization into account),
we obtain a total hydrogen column density of  
$3.45\times 10^{20}$ cm$^{-2}$, corresponding to 
$\Omega_b$(BLA$)=0.0047\,h_{65}\,^{-1}$.
This estimate lies above the original estimate of
$\Omega_b$(BLA$)=0.0025\,h_{65}\,^{-1}$
by Richter et al., suggesting that the consideration
of photoionization is important 
for the estimate of the baryon content of BLAs.
If we restrict our total BLA sample in our simulation to
log $(N/b)\approx 11.3$, log $N$(H\,{\sc ii}$)\leq 20.5$ 
and $b<100$ km\,s$^{-1}$, we obtain
$(d{\cal N}/dz)_{\rm BLA}=18$ and $\Omega_b$(BLA$)=0.0058\,h_{65}\,^{-1}$.
Thus, despite the small-number statistics for BLAs in the
STIS data, observations and simulations are in 
good agreement with each other.

Our study
suggests that the study of BLAs in low-redshift quasar spectra 
represents a useful method to study the distribution and
baryon content of the WHIM.
However, a number of BLA 
candidates significantly larger than currently available is desired to 
constrain the properties of these absorbers on a statistically
secure basis. For this, future FUV measurements of BLAs 
at high spectral resolution and sufficient S/N will be required.

\section{The kinematic relation between BLAs and O\,{\sc vi} absorbers}

A considerable fraction of the BLAs in our sample 
show associated O\,{\sc vi} absorption within
a velocity radius of $50$ km\,s$^{-1}$.
Number distribution and equivalent-width distribution 
of the O\,{\sc vi} absorbers 
are discussed in detail in Fang \& Bryan (2001).
Here we focus on the kinematic relationship between
broad Ly\,$\alpha$ absorption and associated O\,{\sc vi} $\lambda 1031.9$
absorption
in our simulation. Recent STIS observations of BLAs and
associated O\,{\sc vi} absorbers (Richter et al.\,2006) show that 
many of the BLA/O\,{\sc vi} absorber pairs have velocity
offsets between the H\,{\sc i} and the O\,{\sc vi} absorption.
In addition, measured O\,{\sc vi} line widths often are inconsistent 
with those of the broad H\,{\sc i} absorbers (if a single-component 
absorption is assumed). These measurements imply that sub-component
structure is present and/or that the broad Ly\,$\alpha$ absorption
and the O\,{\sc vi} absorption do not trace the same gas phase.
We now want to study these interesting aspects in our artificial
spectra. 

From the 770 BLA systems in our sample, 274 systems
have O\,{\sc vi} absorption that is accurately fitted by
AutoVP. Note that many of these O\,{\sc vi} absorbers 
are composed of several sub-components. For this analysis
we consider only the dominating O\,{\sc vi} sub-component for each
BLA with the highest O\,{\sc vi} optical depth.
In Fig.\,8, upper panel, we have plotted the distribution of
the velocity offsets between the line centroids of the
H\,{\sc i} and the O\,{\sc vi} absorption, 
$\Delta v = |v_{\rm HI} - v_{\rm OVI}|$. O\,{\sc vi} absorbers
that refer to BLAs from the total sample (274 systems) are plotted in gray,
O\,{\sc vi} systems that refer to BLAs in the high-quality sample
(145 systems) are plotted in black.
For the total sample the median
offset between H\,{\sc i} and O\,{\sc vi} is $6.3$ km\,s$^{-1}$,
while for the high-quality sample the median offset is $5.3$ km\,s$^{-1}$.
This implies that small velocity offsets between the H\,{\sc i} Ly\,$\alpha$
absorption and the O\,{\sc vi} absorption are common. 
A possible explanation for such velocity offsets is 
that some
of the gas filaments that
give rise to broad H\,{\sc i} and O\,{\sc vi} absorption 
have a relatively complex (non-Gaussian) density and temperature distribution 
(see, e.g., Fig.\,1, right panel). While the BLAs trace 
the H\,{\sc i} optical-depth weighted density distribution of
the filaments at {\it all} temperatures, O\,{\sc vi} absorbers
preferentially arise from higher-density regions in a small temperature 
range around $3\times 10^5$ K.
Therefore, BLAs and O\,{\sc vi} absorbers 
most likely originate in somewhat different gas phases
and regions within a filament. This would naturally explain the observed small
velocity offsets between these two species in both simulations
and observations.

In the lower panel of Fig.\,8 we compare H\,{\sc i} and O\,{\sc vi}
$b$ values for the total sample (gray dots) and the high-quality
sample (black dots).
If H\,{\sc i} and O\,{\sc vi} would trace the same gas phase
of a purely thermally broadened WHIM absorber one would expect
that $b$(H\,{\sc i}$)=4\,b$(O\,{\sc vi}) owing to the
16 times larger weight of oxygen (see equation (2)).
Most of the BLA/O\,{\sc vi} absorber pairs lie above this
expected relation (in Fig.\,8, lower panel, shown as solid
line) and the data points show a large scatter for both
the total and the high-quality sample. The plot indicates
that the $b$ values of H\,{\sc i} and O\,{\sc vi} are only very poorly
related to each other. This is not surprising in view of the fact
that for most BLAs the absorption of H\,{\sc i} and O\,{\sc vi}
is offset by a few km\,s$^{-1}$ (see above). The O\,{\sc vi} absorbers
have widths significantly larger than what is expected for pure thermal 
broadening in these systems, suggesting that most of the O\,{\sc vi} $b$ values
are determined predominantly by non-thermal processes.

\begin{figure}[t!]
\resizebox{1.0\hsize}{!}{\includegraphics{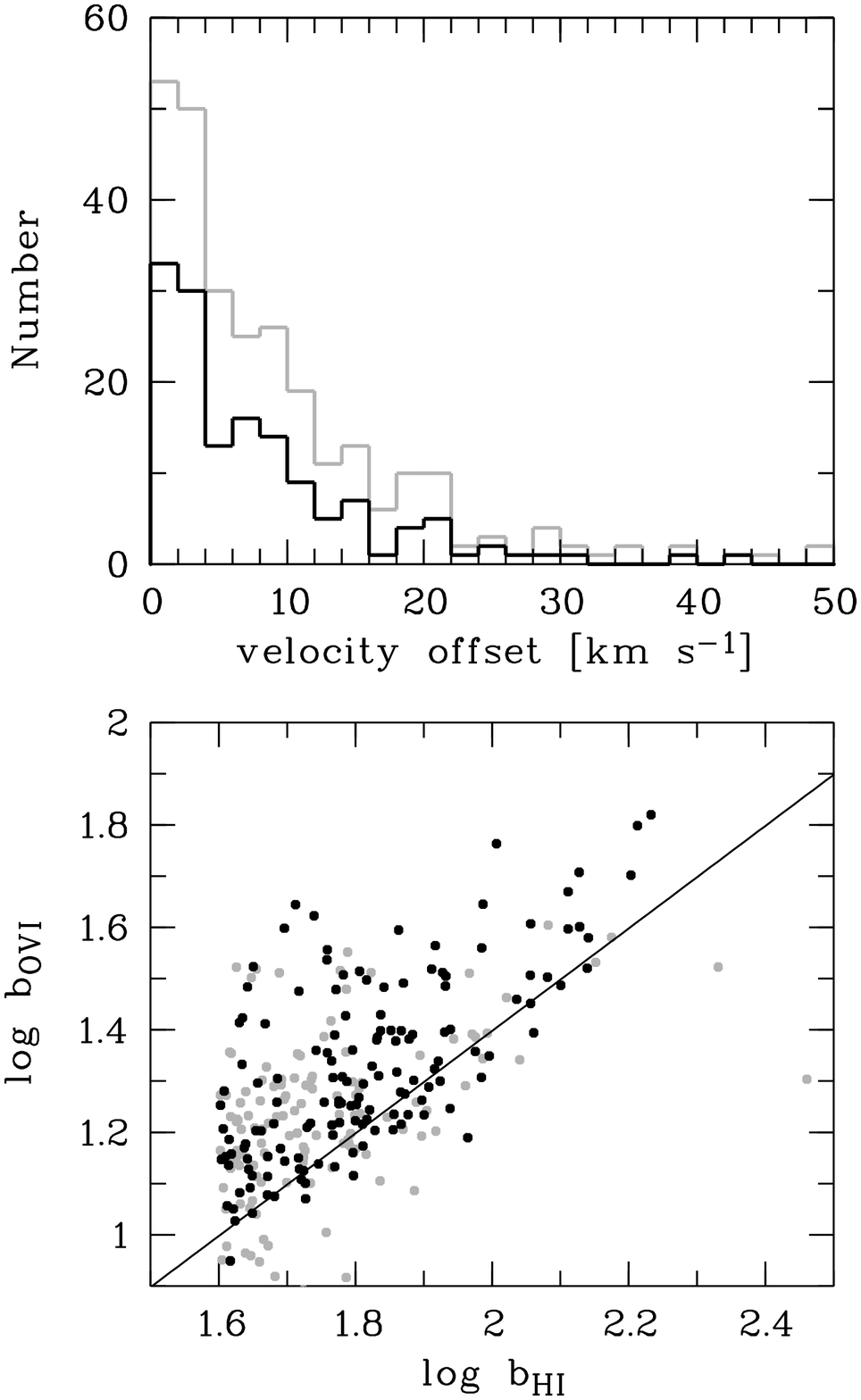}}
\caption[]
{{\it Upper panel:} the distribution of velocity offsets
between broad Ly\,$\alpha$ absorption
and associated O\,{\sc vi} absorption is shown
(gray line = total sample; black line = high-quality sample).
{\it Lower panel:}
O\,{\sc vi} $b$ values are plotted against H\,{\sc i} $b$ values
($b$ in units km\,s$^{-1}$). The
solid line indicates $b$(O\,{\sc vi}$)=b$(H\,{\sc i}$)/4$, as expected
for pure thermal broadening
(gray dots = total sample; black dots = high-quality sample).}
\end{figure}

\section{Summary}

In this paper we have used artificial spectra generated from 
a hydrodynamical simulation to study 
frequency, distribution, physical conditions, and baryon content 
of broad Ly\,$\alpha$ absorbers (BLAs) in the low-redshift 
intergalactic medium. Along a total redshift path of $\Delta z=20$
we find 770 BLA systems with $b\geq 40$ km\,s$^{-1}$
above a detection threshold of    
log $(N$(cm$^{-2})/b$(km\,s$^{-1}))\geq 10.7$, implying a number
density of BLA systems per unit redshift of $(d{\cal N}/dz)_{\rm BLA}=39$
for this sensitivity limit. 
From a spectral analysis with automatted Voigt-profile fitting we
find typical H\,{\sc i} column densities of the 
BLAs of log $N$(H\,{\sc i}$)\approx 13$
and typical $b$ values of $40-100$ km\,s$^{-1}$.
As a visual inspection suggests, only 321 of these BLAs are single-component,
Gaussian shaped absorbers, while the remaining 449 systems show evidence for
sub-component structure and multiple unresolved velocity components. 
Our simulation implies that BLAs predominantly
trace shock-heated gas in the warm-hot intergalactic
medium (WHIM) in the temperature range log $T=4.4-6.2$. However, 
photoionized systems at lower temperatures (log $T<4.3$) also contribute
to the population of BLAs ($\sim 27$ percent by number). 
BLAs that belong to the WHIM typically have overdensities of $20$
and characteristic ionization fractions 
of log $f_{\rm H}\approx 5.6$. At these densities,
both collisional ionization and photoionization contribute to
the total ionization fraction in the gas. We find that 
in the characteristic BLA density range log $f_{\rm H}$ scales 
linearly with log $T$, so that for pure thermal line 
broadening the ionization fraction can be estimated from
the BLA line width. The total hydrogen 
column densities of the BLAs typically lie in the range log 
$N$(H\,{\sc ii}$)=18-20$ with a peak near log 
$N$(H\,{\sc ii}$)=18.8$. Most of the photoionized systems have
logarithmic total hydrogen column densities $<18.5$ and thus
do not contribute significantly to the the total baryon content of BLAs.
We find that non-thermal line broadening mechanisms contribute (on average) with
$\sim 10$ percent to the total $b$ value of a WHIM BLA, suggesting that
for most systems the measured line width represents a good 
indicator for the thermal state of the absorbing gas. Therefore, our
simulation suggests that
it is possible to estimate the total gas column density in BLAs 
from the measured $b$ values and H\,{\sc i} column densities and
to estimate the baryon content of these systems 
in observational data. Our data imply that this indirect method
works well for systems with total logarithmic hydrogen column densities
$<20.5$ and for a BLA sample that is sufficiently large. 
The total baryon content of all BLAs in our simulation 
in the range log $N$(H\,{\sc ii}$)\leq 20.5$, $b\geq 40$ km\,s$^{-1}$ and
log $(N/b)\geq 10.7$ (754 systems; $d{\cal N}/dz\approx 38$) 
is $\Omega_{b}$(BLA$)=0.0121\,h_{65}\,^{-1}$, which
represents $\sim 25$ percent of the total baryon budget
in our simulation.
This result supports the idea that BLAs represent
a significant baryon reservoir at low redshift. If we restrict
our total BLA sample to the range log $N$(H\,{\sc ii}$)\leq 20.5$, 
$b=40-100$ km\,s$^{-1}$ and log $(N/b)\geq 11.3$,
we obtain $(d{\cal N}/dz)_{\rm BLA}\approx 18$ 
and $\Omega_{b}$(BLA$)=0.0058\,h_{65}\,^{-1}$, in
agreement with recent FUV observations.

Our simulation indicates that future observational studies of 
BLAs at low $z$ will be of great importance to study the baryon
content of the WHIM, but for a reliable estimate of 
$\Omega_{b}$(BLA) a large sample ($n>10$) of high-resolution, 
high S/N FUV spectra of low redshift QSOs will be required. 
The {\it Cosmic Origin Spectrograph} (COS), originally scheduled 
to be implemented on the Hubble Space Telescope (HST) in 2004, 
would have been the ideal instrument for these 
observations, but in view of the uncertain status of the 
COS mission and other future FUV instruments it remains
unclear whether such data will be available in the near future.
Next to observations, additional simulations of
the low-redshift IGM with higher resolution and more realistic
physics (e.g., galaxy feedback, metal-line cooling, non-equilibrium
processes, etc.) will be important to investigate the physical conditions of 
BLAs in more detail and to re-examine the results presented in this paper.

\begin{acknowledgements}

P.R. acknowledges financial support by the German
\emph{Deut\-sche For\-schungs\-ge\-mein\-schaft}, DFG,
through Emmy-Noether grant Ri 1124/3-1. 
T.F. is supported by the NASA through {\sl Chandra} Postdoctoral Fellowship
Award Number PF3-40030 issued by the {\sl Chandra} X-ray Observatory
Center, which is operated by the Smithsonian Astrophysical Observatory for
and on behalf of the NASA under contract NAS 8-39073.
G.L.B. is partially supported by NSF grant AST0507161. We thank
Chris McKee for helpful comments.

\end{acknowledgements}

{}

\end{document}